\documentclass{emulateapj}
\newcommand{\be}{\begin{equation}}
\newcommand{\ba}{\begin{eqnarray}}
\newcommand{\ee}{\end{equation}}
\newcommand{\ea}{\end{eqnarray}}
\newcommand{\mumag}{\mu{\rm mag}}
\newcommand{\ga} {\gtrsim}
\newcommand{\la} {\lesssim}

\begin{document}

\journalinfo{The Astrophysical Journal, 589, L000, 2003 May 10}

\lefthead{LOEB \& GAUDI} 
\righthead{PERIODIC DOPPLER FLUX VARIABILITY}

\title{Periodic Flux Variability of Stars due to the \\
Reflex Doppler Effect  Induced by Planetary Companions}

\author{Abraham Loeb \altaffilmark{1} and B.\ Scott Gaudi\altaffilmark{2}}

\affil{Institute for Advanced Study, Princeton, NJ 08540}
\email{loeb@ias.edu,gaudi@ias.edu}

\altaffiltext{1}{Guggenheim Fellow; on sabbatical leave from the
Astronomy Department, Harvard University, Cambridge, MA 02138}

\altaffiltext{2}{Hubble Fellow}

\begin{abstract}
Upcoming space-based photometric satellites offer the possibility of
detecting continuum flux variability at the micro-magnitude ($\mu{\rm mag}$)
level.  We show that the Doppler flux variability induced by the
reflex motion of stars due to planetary companions has an amplitude of
$(3-\alpha)K/c$, where $K$ is the reflex radial velocity amplitude and
$\alpha\approx d \ln F_\nu/d\ln \nu$ is the logarithmic slope of the
source spectral flux in the observed frequency band.  For many of the
known close-in planetary systems with periods $P\la 0.2~{\rm yr}$, the
periodic Doppler variability, ${\cal O}(\mumag)$, is significant
relative to the variability caused by reflected light from the
planetary companion. For companions with $P\ga 0.2~{\rm yr}$, the
Doppler signal is larger than the reflected light signal.  We show
that the future photometric satellites should reach the sensitivity to
detect this Doppler variability.  In particular, the {\it Kepler}
satellite should have the photon noise sensitivity to detect at a
signal-to-noise ratio $S/N \ga 5$, all planets with minimum mass $M_p
\sin i \ga 5M_J$ and $P\la 0.1~{\rm yr}$ around the $\sim 10^4$
main-sequence stars with spectral types $A$ through $K$ and apparent
magnitude $V<12$ in its field-of-view.

\end{abstract}


\section{Introduction}

Planets outside the solar system orbiting main-sequence stars
were first discovered in the mid-1990s 
(Mayor \& Queloz 1995; Marcy \& Butler 1996) through the
small periodic reflex motion they imprint on the radial velocity of
their primary stars. The planetary systems discovered by the radial
velocity surveys to date are qualitatively different from the Solar 
System in important ways that were unexpected by earlier theoretical 
models of planet formation.

Although radial velocity studies have been highly effective at
identifying more than a hundred extrasolar planetary
systems\footnote{See http://cfa-www.harvard.edu/planets/catalog, for a
list of planets and discovery references.}, the method is inherently
limited in several ways.  First, it can generally only be used to find
planets orbiting old, late-type stars.  Second, it is not sensitive to
planets with mass $M_p\la M_\oplus$.  Finally, one can only infer the
semi-major axis $a$, eccentricity, and a lower-limit to the planet
mass.  It is therefore important to seek new, complementary techniques
that will allow better characterization of the physical properties of
the known planets, and extend the current sample to different types of
stars and lower-mass planets.

The power of combining the radial velocity method with other
techniques was best illustrated in the case of the planetary companion
to HD209458.  Originally discovered using radial velocity surveys, this
planet was subsequently found to transit its parent star.  Combination
of the radial velocity data with the original ground-based transit
data allowed the measurement of the mass (without an inclination angle
ambiguity), radius and density of this planet (Charbonneau et al.\
2000; Henry et al.\ 2000).  Further ultra-high precision HST/STIS
observations of several transits allowed the exquisite measurement of
these planetary properties, as well as the detection of sodium lines
in its atmosphere (Brown et al. 2001; Charbonneau et al. 2002).
Perhaps the most exciting aspect of the HST measurements was the
demonstration of the ability to achieve photometric precision at the
level of $\sim 10^{-4}$ with space-based observations.

In the future, several space-based photometric satellite missions are
expected to reach a precision better than that demonstrated for
HD209458, down to the micro-magnitude ($\mumag$) level.  These
include {\it Microvariability and Oscillations of STars
(MOST)}\footnote{http://www.astro.ubc.ca/MOST/}, {\it COnvection,
ROtation and planetary Transits
(COROT)}\footnote{http://smsc.cnes.fr/COROT/}, {\it
Kepler}\footnote{http://www.kepler.arc.nasa.gov/}, and {\it
Eddington}\footnote{http://sci.esa.int/home/eddington/index.cfm}.
Scheduled for launch in June 2003, {\it MOST} is a targeted mission,
and among its primary targets are five stars with known extrasolar
planetary companions.  {\it COROT}, {\it Kepler}, and {\it Eddington},
will monitor thousands of stars in one or several fields and
search for very low-amplitude variability, as produced by stellar
oscillations and planetary transits.

At the $\mumag$ levels of variability probed by these satellites, many
otherwise undetectable effects associated with planetary companions
become significant.  For transiting planets, the effects of the
quadrupole moment of the parent star and more distant companions
(Miralda-Escud{\' e} 2002), the oblateness of the planet (Hui \&
Seager 2000, Seager \& Hui 2002), and the presence of moons and rings
associated with the planet (Sartoretti \& Schneider 1999, Brown et
al. 2001), are all potentially detectable.
For both transiting and non-transiting companions, flux variations
arising from tidal distortion of the parent star due to the planet
(von Zeipel 1924, Drake 2003), back-heating of the star from the
planet (Green et al.\ 2003), and reflected and scattered light from
the planet's atmosphere (Seager, Whitney, \& Sasselov 2000; Sudarsky,
Burrows, \& Hubeny 2003; Green et al.\ 2003), may all be significant.

Here we examine another potential source of periodic flux variability,
namely the photometric flux variation of the primary star due to the
{\it Doppler effect}. The reflex motion of the star induced by its
planetary companion, causes slight Doppler variations in its
photometric flux. In \S 2 we derive the amplitude of this effect.  In
\S 3 we compare the Doppler effect with other sources of stellar flux
variability in the presence of planetary companions.  In \S 4
we demonstrate that the Doppler variability should be detectable with
the upcoming {\it MOST} and {\it Kepler} missions.  Finally, we discuss
the systematic errors and intrinsic variability in \S 5, and summarize
our conclusions in \S 6.

\section{Doppler Flux Variations}

A star moving at a radial {\it non-relativistic} velocity $v_r$
relative to the observer, obtains a Doppler shift to its bolometric
flux by an amount [see Eq. (4.97b) in Rybicki \& Lightman 1979], 
\be
F= F_0\left(1+4{v_r\over c}\right), 
\ee 
where $c$ is the speed of light and the
subscript $0$ denotes observed quantities in the absence of the source
motion.  If the emitted flux per unit frequency scales as a power-law
in frequency, $F_{\nu 0}\propto \nu^{\alpha}$, then the Doppler
shift in the observed frequency $\nu=\nu_0(1+ v_r/c)$ implies
that the observer detects, 
\be 
F_\nu= F_{\nu 0}\left[1+
(3-\alpha){v_r\over c}\right] .
\label{eq:spec_flux}
\ee 
The Doppler effect vanishes for $\alpha=3$ because $F_\nu/\nu^3$
is a relativistic invariant (corresponding to the phase space density
of photons). Therefore, $F_\nu=F_{\nu 0}$ for $\alpha=3$ irrespective
of the source motion.  Any other value of $\alpha$ leads to a net
Doppler effect due to the change in the photon number flux.

For a general source spectrum one may substitute $\alpha\approx (d \ln
F_{\nu 0}/d\ln \nu_0)$, where the derivative is averaged over the
observed band of frequencies.  Assuming a blackbody source spectrum
with a temperature $T_{\rm eff}$, \be \alpha(\nu) \simeq
\frac{e^{x}(3-x)-3}{e^{x}-1}, \ee where $x\equiv {h\nu/kT_{\rm eff}}$.
In the Rayleigh-Jeans part of the spectrum ($x \ll 1$), $\alpha=2$,
whereas in the Wien tail ($x\gg 1$), $\alpha=3-x$.  For solar-type
stars ($T_{\rm eff}\simeq 5700~{\rm K}$), and observations in the
optical band ($\lambda\approx 600~{\rm nm}$ or $\nu \approx 5 \times
10^{14}~{\rm Hz}$), we get $\alpha \approx -1.3$.  Given that the
amplitude of the Doppler signal and the amount of flux received are
both functions of frequency, the signal-to-noise ratio will also
depend on the observed frequency.  For Poisson errors, the noise
variance will be proportional to the square-root of the flux.  We find
that the signal-to-noise ratio for $T_{\rm eff}\sim 5700~{\rm K}$ and
fixed logarithmic frequency interval peaks in the $V$-band, at
$\lambda \sim 500~{\rm nm}$ or $\nu \sim 6 \times 10^{14}~{\rm Hz}$.

\section{Application to Planetary Companions}
 
The reflex motion of the star due to a planetary companion should
produce periodic variations in its observed 
\begin{figure}[htbp]
\plotone{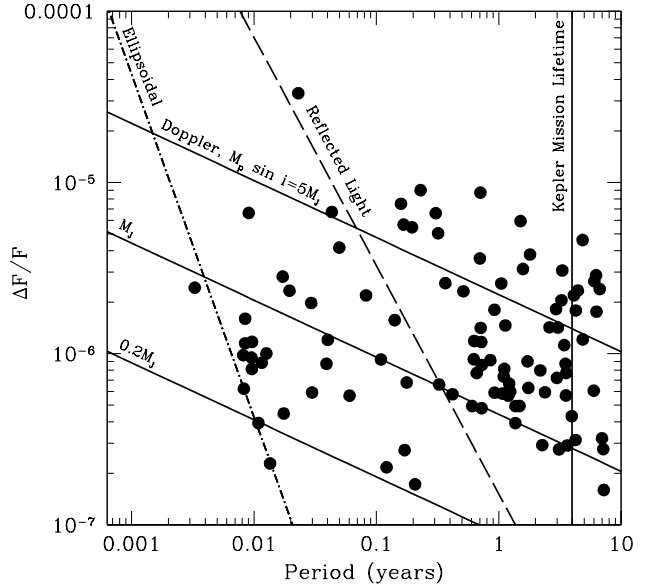}
\caption{
Expected flux variation amplitude as a function of the orbital period
for the known extrasolar giant planets (filled dots).  
Also shown are the expected signals for fixed
planet masses, assuming circular orbits, inclinations of 90$^\circ$, and
a solar-type primary.  The dashed line
shows the level of expected reflected light variations for a
Jupiter-radius planet, with Lambert sphere scattering (albedo = 2/3)
and at full phase.  The dashed-dot line shows
the level of expected ellipsoidal light variations 
for a Jupiter-mass planet and solar-type primary.  
The solid vertical line is the Kepler mission lifetime (4 years). 
}
\label{fig-tau1}
\end{figure}
spectral flux.  For
circular orbits, the radial velocity of the star relative to the
observer is simply $v_r=K\sin 2\pi\phi$, where $\phi$ is the phase of
the orbit, and $K$ is the velocity semi-amplitude, 
\be K=28.4~{\rm
m~s^{-1}}\left(\frac{P}{1~{\rm yr}}\right)^{-1/3}\left(\frac{M_p\sin
i}{M_J}\right) \left(\frac{M_*}{M_\odot}\right)^{-2/3}.
\label{eq:kamp}
\ee 
Here $M_p$ is the mass of the planet, $M_J$ is the mass of
Jupiter, $i$ is the inclination of the orbit, $M_*$ is the mass of the
star, and we have assumed that $M_p \ll M_*$. Thus, the flux variability
signal has a period $P$ and fractional amplitude $\Delta
F/F_0=(3-\alpha)K/c$.  For orbits with an eccentricity $e$, the
right-hand-side of equation~(\ref{eq:kamp}) should be multiplied by
a factor $(1-e^2)^{-1/2}$, which is typically very close to unity.

Since $K$ is a direct observable for companions detected
spectroscopically via radial velocity variations, we can predict the
amplitude of the the flux variations for known planetary companions
with essentially no ambiguity.  Figure 1 shows the expected flux
variation amplitude as a function of the orbital period for all the
known radial-velocity planets that have been detected so far,
including the new transiting planet OGLE-TR-56 (Udalski et~al.\ 2002,
Konacki et~al.\ 2003).  We have estimated $T_{\rm eff}$ values from
the known spectral types of the host stars, and assumed observations
in the $V$-band.  Also shown are the expected signals as a function of
period for fixed planet masses, assuming circular orbits, inclination
of 90$^\circ$, solar-type primaries, and observations in the $V$-band.

The reflected light amplitude for a planet of radius $R_p$ is
\be
\frac{\Delta F}{F_0}=p \left(\frac{R_p}{a}\right)^2.
\label{eqn:epsilon}
\ee 
Here $p$ and $R_p$ are the geometric albedo and the radius of the
planet, and $a$ is the semi-major axis of its orbit.  The dashed line
shows $\Delta F/F_0$ for Lambert sphere scattering ($p=2/3$), and
planets at full phase.  Averaged over all phase angles, Lambert sphere
scattering represents the upper limit to the amount of reflected light
for a wide range of reasonable models, although larger scattered light
signals are possible near zero phase angle (when the planet is at
superior conjunction) due to, e.g.\ strong back-scattering (Seager
et~al.\ 2000).

For short-period ($P\la
0.2~{\rm yr}$), massive ($M_p\sin i\ga M_J$) planets, the Doppler
variability is expected to be a significant contaminant to the
reflected light signal.  For known radial-velocity companions, the
Doppler effect is completely predictable and its subtraction is
straightforward.  When searching for planets via reflected
light, the magnitude of the Doppler effect will not be known a priori.
However, since the reflected light signal will generally be
largest when the planet is at superior conjunction, it will be $90^\circ$ out
of phase with the Doppler signal, and it should be possible to
cleanly disentangle the two signals.  The Doppler signal
is expected to dominate for long periods of $P\ga 0.2~{\rm yr}$, since
the reflected light signal falls off more rapidly with increasing
orbital radius ($\propto a^{-2}$) than the Doppler signal ($\propto
a^{-1/2}$).

There will also exist ellipsoidal light variations due to tidal
effects on the star from the planet.  We estimate the amplitude of
these variations as, 
\be 
\frac{\Delta F}{F_0}\sim \beta
\frac{M_p}{M_*}\left(\frac{R_*}{a}\right)^3.  
\ee 
where $R_*$ is the radius of the primary, and $\beta\simeq 0.45$ is
the `gravity-darkening' exponent for solar-type stars.  The amplitude
of the expected ellipsoidal variations as a function of period is
shown in Figure 1 as the dot-dashed line for $M_p=M_J$, $M_*=M_\odot$,
and $R_*=R_\odot$.  This effect should be subdominant for all but the
shortest periods.  Green et~al.\ (2003) predict that back-heating of
the star by the planet should be an even smaller effect, ${\cal
O}(10^{-7})$.

\section{Detectability}

For observations over a net duration $T$, the expected
signal-to-noise ratio can be estimated as, 
\be 
\frac{S}{N} =
\frac{(3-\alpha)}{\sqrt{2}}\frac{K}{c} \sqrt{\Gamma \times~T\times~
10^{-0.4V}},
\label{eq:sn}
\ee where $\Gamma$ is the number of photons per unit time collected by
the instrument for a star of magnitude $V=0$, and we have assumed
$P\ll T$.  Note that the $\sqrt{2}$ factor is just the
root-mean-square of the sinusoidal signal (for a planet in a circular
orbit) in units of the amplitude.  One can improve the signal-to-noise
ratio by $\sqrt{2}$ if the period and phase of the signal are known in
advance by making observations when the signal is at maximum.

\begin{figure}[htbp]
\plotone{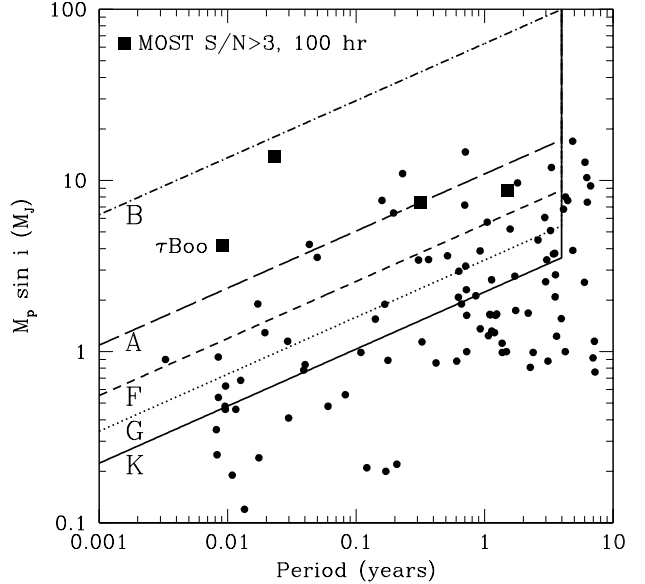}
\caption{ Mass as a function of orbital period for the known
extrasolar giant planets (filled dots).  The large solid squares
show those known planets for which the signal-to-noise ratio ($S/N$) of the
flux variation is $>3$  for 100 hours of integration time
with the {\it MOST} satellite.  
The lines show the $S/N>5$ detection limits for the {\it Kepler}
mission for stars with
$V=12$, and spectral types B, A, F, G, and K (upper to
lower diagonal lines).}
\label{fig-tau2}
\end{figure}

The {\it MOST} satellite will collect $\Gamma \sim 1.6\times 10^8$
photons per second for a $V=0$ star (Green et~al.\ 2003).  In Figure
2, we show $M_p\sin i$ versus period of the known extrasolar planets.
Taking into account the $V$-magnitudes of the parent stars, we have
calculated the expected signal-to-noise ratio $(S/N)$ for a total of
100~{\rm hours} of integration with {\it MOST}.  Since {\it MOST} is a
targeted mission, we have assumed that the observations can be spaced
over many periods and made at phases when the signal is largest. Thus
we assume a signal-to-noise ratio that is a factor of $\sqrt{2}$
larger than given in equation (\ref{eq:sn}).  We have highlighted
those stars for which $S/N\ge 3$.  Of these, one is currently included
in the {\it MOST} target list: $\tau$ Boo, for which $S/N\sim 6$ is
expected.

The {\it Kepler} satellite will look at a single field continuously
for $T=4$ years, and will collect $\Gamma \simeq 1.4\times 10^{10}$
photons per second for a $V=0$ star.  The detection limits for
primaries of spectral type B-K are shown in Figure 2, assuming a total
integration time of $T=4$ years, $V=12$, circular, edge-on orbits, and
$S/N\ge 5$ for detection.  There are $\sim 10^4$ such main-sequence
stars in {\it Kepler}'s field-of-view.  If the frequency of massive
($M\ga 5M_J$), short-period ($P\la 0.1~{\rm yr}$) planets is 0.1\%,
{\it Kepler} should detect $\sim 10$ such companions.

Both {\it MOST} and {\it Kepler} will have broad wavelength filters in
order to gather as many photons as possible.  {\it MOST}'s bandpass
will cover $350$--$700~{\rm nm}$, whereas {\it Kepler}'s bandpass will
cover $400$--$850~{\rm nm}$.  The spectra of the sources will vary
considerably over these wavelength ranges, and our assumption of a
fixed spectral slope appropriate to the center of the $V$-band will
break down.  We have estimated the magnitude of this effect by
integrating the blackbody spectrum over a uniform bandpass from
$400$--$800~{\rm nm}$. We find that the expected signal-to-noise ratio
is smaller by $\sim 5\%$ than we have estimated before.  The source
spectra will deviate from the blackbody shape due to the presence of,
for example, spectral lines.  To determine the effects of such
deviations, we have calculated the signal-to-noise ratio using a
high-resolution Solar spectrum of Kurucz et~al.\ (1984), kindly
provided to us by E.\ Turner.  Again, we find that the difference from
the naive calculation is small, $\sim 6\%$.

\section{Systematic Errors and Intrinsic Variability}

The amplitudes of the Doppler effects we have considered are extremely
small, even for the extremely precise upcoming space-based photometric
missions.  We therefore briefly consider both systematic errors and
intrinsic variability.  

The photometric stability requirements for the detection of the
signals we consider are considerably more stringent than the
requirements for the detection of the signals these satellites are
designed to measure.  To achieve the photon limit during the duration
of the transit of an Earth-like planet, the photometry must be stable
to roughly 1 part in $10^5$ over the $\sim 10~{\rm hour}$ duration of
the transit.  In contrast, we require that the photometry be stable to
roughly 1 part in $10^6$ at least over the $\sim 3~{\rm day}$ periods
of the smallest planetary orbits.  Reflected light signals also
require stability over similar time scales, but in the cases where
these signals are several times larger, the stability requirements
are less severe.

Not much is known about intrinsic variability at the levels we are
considering for any stars except for the Sun.  The intrinsic
variability of the Sun is $\la 10^{-5}$ on time scales of a few days
or less (Batalha et~al.\ 2001).  This is of order the expected
integrated Poisson error over these time scales, and thus should not
significantly affect our signal-to-noise ratio estimates.  However, the Sun
does show significant, of order a few parts in $10^{4}$, variability on
30-day periods or longer (Batalha et~al.\ 2001).  This variability is
a consequence of magnetic activity and sunspots.  Depending on the
coherence of its periodicity, this variability may complicate the
detection of planets with longer periods via the Doppler
variability.  If measurements in two frequency bands are available, it
may be possible to isolate the Doppler effect through its unique
dependence on $\alpha$ (see Eq.~\ref{eq:spec_flux}).

\section{Conclusions}

Equation (\ref{eq:spec_flux}) provides the amplitude of the Doppler
flux variability induced by the reflex motion of stars due to close-in
planetary companions.  For many of the known close-in planetary
systems, the Doppler variability will be similar or larger than the
variability due to reflected light from the planetary companion (see
Fig.\ 1).  

We have found that future space-based photometry missions should reach
the required photon-noise sensitivity for the detection of this
Doppler effect.  In particular, {\it MOST} should detect this effect
due to the known companion of $\tau$ Boo at a signal-to-noise ratio 
of $\sim
6$ in 100 hours of integration.  {\it Kepler} should detect all
planetary companions with masses $\ga 5M_J$ and periods $\la 0.1~{\rm
yr}$ to the $\sim 10^4$ main-sequence stars with $V<12$ in its
field-of-view (see Fig. 2).  This may uncover planets around massive
primaries that are inaccessible to radial velocity surveys and too
distant from their primary stars to be detected via reflected light.

\acknowledgements

We thank Andy Gould, Jon Jenkins, Sara Seager, Ed Turner, Josh Winn,
and the referee, Ron Gilliland, for useful discussions and comments.
A.L. acknowledges support from the Institute for Advanced Study at
Princeton and the John Simon Guggenheim Memorial Fellowship.  This
work was also supported in part by NSF grants AST-0071019, AST-0204514
and NASA grant ATP02-0004-0093 (for A.L.), and in part by NASA through
a Hubble Fellowship grant from the Space Telescope Science Institute,
which is operated by the Association of Universities for Research in
Astronomy, Inc., under NASA contract NAS5-26555 (for S.G.).  The
NSO/Kitt Peak FTS data used here were produced by NSF/NOAO.

\end{document}